# Decoding multimodal behavior using time differences of MEG events

Ohad Felsenstein[1], Idan Tal[1,2], Michal Ben-Shachar[1,3], Moshe Abeles[1,4], Gal Chechik[1,5]

[1] Gonda Multidisciplinary Brain Research Center, Bar-Ilan University, Ramat-Gan, Israel.
[2] Nathan Kline institute for psychiatric research Orangeburg, NY and Columbia University medical center, department of neurological surgery, NY.
[3] Department of English Literature and Linguistics, Bar Ilan University, Ramat Gan, Israel.
[4] The Hebrew University of Jerusalem, Jerusalem, Israel.
[5] NVIDIA Research.

## Abstract

Multimodal behavior involves multiple processing stations distributed across distant brain regions, but our understanding of how such distributed processing is coordinated in the brain is limited. Here we take a decoding approach to this problem, aiming to quantify how temporal aspects of brain-wide neural activity may be used to infer specific multimodal behaviors. Using high temporal resolution measurements by MEG, we detect bursts of activity from hundreds of locations across the surface of the brain at millisecond resolution. We then compare decoding using three characteristics of neural activity bursts, decoding with event counts, with latencies and with time differences between pairs of events. Training decoders in this regime is particularly challenging because the number of samples is smaller by orders of magnitude than the input dimensionality. We develop a new decoding approach for this regime that combines non-parametric modelling with aggressive feature selection. Surprisingly, we find that decoding using time-differences, based on thousands of region pairs, is significantly more accurate than using other activity characteristics, reaching 90% accuracy consistently across subjects. These results suggest that relevant information about multimodal brain function is provided by subtle time differences across remote brain areas.

## Author Summary

We quantified inference of multimodal behavior from MEG activity events using three characteristics of neural activity, using spatial, spatio-temporal and time-difference information. Behavior can be inferred with 90% accuracy using time-differences of activity bursts (cortical events) distributed across the brain. This decoding scheme achieved significantly more accurate performance compared to other characteristics of neural activity, using the same evaluation framework, in every subject tested. Interestingly, some of the activity patterns are consistent across subjects. These results are consistent with the hypothesis that high-level processes involve a network of numerous processing stages, each eliciting timed activity bursts across the brain.

## 1. Introduction

Many daily tasks, from walking down a busy street to typing a letter, combine processes across multiple modalities. To process and coordinate such multimodal tasks, the neural system has to coordinate activity across numerous, possibly remote, brain regions. Unfortunately, our understanding of the neural mechanisms underlying such coordination is still limited.

This coordination problem has a long history in neuroscience. One early proposal suggests that integration of multiple processes is achieved by synchronizing their corresponding neuronal activity [1]. This idea is supported by studies reporting that groups of spatially-distributed cortical neurons are involved in synchronous activity [2–9]. These findings suggest that the cerebral cortex uses the time domain to code information, by synchronizing neuronal discharges with millisecond precision [10]. It was further hypothesized that rhythmic synchronization can dynamically tune the effective connectivity *(Communication through coherence* [3,11]). By doing so, it enables communication between distant neuronal groups with high temporal precision[12–15].

Going beyond synchrony, several studies found systematic activity **time lags** between groups of neurons that communicate with each other[16–19]. Such consistent time lags may reflect collaborative information processing and carry information about the functional connectivity of neurons that show systematic time lags. More specifically, time lags in neural activation can carry information about the order of neuronal events within a functional pathway. Indeed, it has long been proposed that activity in neural systems can be viewed as a succession of transient events (see recent reviews by [20,21]). For instance visual processing can be viewed as cascading from the retina and lateral geniculate nucleus, through primary visual cortex, and along the ventral and dorsal streams, and similarly for perception in general [e.g. 22–25].

The idea of processing cascades also applies to tasks that involve multiple modalities, for instance, in a task called ***sensorimotor synchronization*** (SMS). SMS is the coordination of rhythmic movements with an external rhythm, which applies to finger tapping in synchrony with a metronome or to musical ensemble performance [26]. To perform well in SMS, the brain needs to coordinate multiple systems, including auditory input and rhythm perception, timing, motor systems, feedback and error correction (for review see 28).

If brain activity reflects sequential processing of information along several parallel streams, one may view neural activity as a sequence of well-timed events, possibly with timing noise. According to this view, timing noise accumulates along the processing pathway, making it easier to characterize subsequent events using their *relative timing* rather than using their spatial distribution or absolute timing [28]. **We hypothesize that pairwise time differences of cortical responses are a key aspect of neural activity during integrative multimodal behavior**. A direct prediction of this hypothesis is that such time differences provide significant information about multimodal behavior, beyond that provided by event counts or event latencies referenced to stimulus onset. As far as we know, this idea has not been systemically explored before in the analysis of human brain imaging data.

To test this hypothesis quantitatively, we take a *decoding* approach and quantify how well behavior can be inferred from pairwise timing differences compared with counts and latencies of the same events. While successful decoding does not guaranty that the features used for decoding are the ones used by the brain, successful decoding gives a lower bound to the amount of information carried by the model features [29–31]. Such analysis relies on capturing spatio-temporal patterns across non-neighboring regions, and therefore requires to (1) use a behavioral task that elicits an extensive pattern of brain activity, and (2) record neural activation across the entire brain at high spatio-temporal resolution. To achieve this, we chose here to use Magnetoencephalography (MEG), a non-invasive technique that has millisecond temporal resolution and a fair spatial resolution, with whole brain coverage.

Significant progress was made recently in developing decoding methods based on MEG recordings from single-trials [32,33]. Several decoding methods focused on finding better input representations and selecting features [34–37], while others focused on reducing data dimensionality [38], developing new decoding approaches [39], and improving model interpretability [40–43]. Another thrust of MEG decoding research focused on applicative aspects, including developing brain-machine interface for motor control [44,45], probing attention in a scene with multiple speakers [46] and measuring endogenous pain [47].

Additionally, MEG decoding has been instrumental for gaining new insights about neural processing and representations. MEG was used to decode abstract visual patterns [48], to infer objects in an object recognition task [49,50], recognizing emotion in faces [51], study processes underling different auditory novelty detection types [24], and to decode objects embedded in various scenes [52]. Broadly speaking, these previous studies make use of the high temporal precision of MEG to extract features in the frequency domain, to identify when processing occurred, or to decode the state of a subject as it evolves over time. In contrast with these approaches, the current paper tests the utility of timing differences between spatially distributed MEG events in decoding multimodal behavior.

We systematically tested three types of neural activity signatures as features for decoding: *spatial decoding* - based on the spatial distribution of activity bursts across the brain, *spatio-temporal decoding* - based on latencies of activity bursts, and *time-difference decoding* - based on time differences between pairs of activity bursts across the brain. We compared the amount of information that is carried by the different decoding schemes by comparing the performance of the same decoder using three different features.

Inferring behavior from MEG recordings is hard because the space of possible activity combinations is immense, and the number of samples collected in a typical experiment is very limited. Together, these issues pose a difficult challenge for developing a robust inference method. Furthermore, it is often difficult to gain insight from "black-box" decoding approaches, especially with high input dimensions. To make the decoder more insightful and interpretable, we design it to be sparse both in space- localizing activity to brain regions, and in time – detecting time-localized events of neural activity. This sparsity means that we can identify a relatively small number of location pairs together with their event timing differences, allowing us to gain some insight into their possible function.

## 2. Materials and methods

**Participants**

Eight volunteers (5 female) participated in the experiment and received financial reimbursement for their time and travel expenses. All subjects signed an informed consent form prior to their participation in the study. The protocol was approved by the ethics committee of Bar-Ilan University. All subjects were right handed (by self-report) and had no more than two years of formal musical training.

**The behavioral task**

Our goal is to compare decoding schemes in neural activity underlying rich multimodal behavior. We adapted a commonly-used paradigm from the field of SMS, where subjects are asked to tap in synchrony with an external stimulus (see reviews by 25,51). The task used here is a variant of SMS tapping paradigm which also includes the perception of musical meter [28]. We expected this task to elicit rich patterns of activity across the brain, activating networks of information processing regions.

Fig 1 illustrates the task. Subjects were asked to listen to a sequence of drum beats that generated a certain meter (e.g., 3/4), and to tap their fingers along with the beat while using their index finger for the accented (primary) beats and middle finger for unaccented (secondary) beats. At random points in time, the meter changed, and the subject had to adjust his tapping accordingly [28]. The decoder goal is to classify activity patterns into one of two conditions: (1) *Before-change* - a synchronous tap (see two examples marked in red in Fig 1); (2) *After-change* – the first asynchronous tap (see examples marked in blue).

More specifically, subjects listened to drum beats with a double meter, where the primary beat was followed by *one* secondary beat, or triple meter where the primary beat is followed by *two* secondary beats. The beat-beat interval was fixed at 0.493s. A single trial spanned 0 to 0.490s from the stimulus onset. We collected a similar number of trials for the *before-change* and *after-change* conditions, ~91 trials for each. The actual number of trials per condition varied in 86–94, depending on the cleaning procedure of individual participants. Importantly, trials in the *after-change* condition contained equal number of trials from two types of incongruency: hearing a primary or secondary beat and tap with the middle or index finger respectively. Similarly, trials in the *before-change* condition contained both types of congruent tap (hearing primary beat and tap with the index finger or hearing secondary beat and tap with the middle finger). This balanced setup was designed to ensure that the decoding captures a higher cognitive process rather than simply learn to discriminate between the sounds or motor actions.

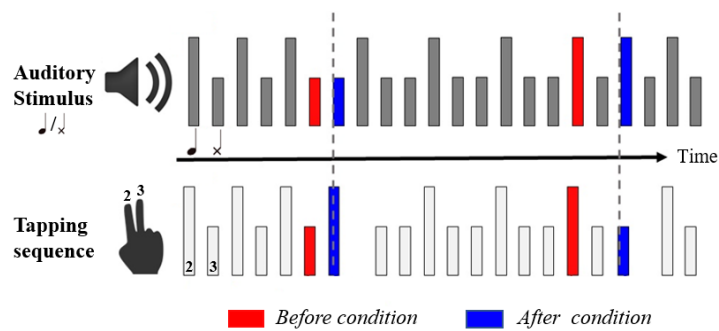

**Fig 1. The behavioral task**. **Top:** The auditory stimulus consists of a sequence of drum beats: Tall bars indicate primary beats and short bars indicate secondary beats. **Bottom:** The tapping sequence. Tall bars indicate tapping with the 2$^{nd}$ finger, short bars indicate tapping with the 3$^{rd}$ finger. Red shading indicates the "before" condition, immediately preceding a change in meter. Blue

shading marks the "after" condition, immediately following a change in the meter of the auditory stimulus.

**MEG data acquisition**

This paper is a re-analysis of a subset of the subjects that were recorded at high sampling frequency of 1017.25Hz. For more details about the data acquisition see [28]. In short, MEG data were acquired using a whole-head helmet-shaped biomagnetometer (4D-Neuroimaging, San Diego). The sensor array consisted of 248 superconducting magnetometers. A head-position indicator using five coils attached to the scalp provided information on head position relative to the sensor array, before and after the measurement. Coil position were determined based on external anatomical landmarks (left preauricular, right preauricular and nasion). The head shape and coil positions were digitized using a Pollhemus FASTTRAK digitizer. MEG signals were band-pass filtered online at 0.1-400Hz.

**MRI data acquisition**

Magnetic Resonance Imaging (MRI) data were collected using a 3T scanner (Signa Excite, General Electric Medical Systems, Milwaukee, WI) located at the Tel Aviv Sourasky Medical Center. High resolution T1 anatomical images were acquired for each participant using a 3D fast spoiled gradient-recalled echo sequence (FSPGR; 150± 12 1-mm thick axial slices, covering the entire cerebrum; voxel size: 1×1×1 mm). For more details, see [54].

**MEG data preprocessing**

MEG recordings were cleaned for power-line frequency, heartbeat artifacts and 24Hz building vibration artifact. For more details about the cleaning process, see [55]. Segments containing eye movements and eye blinks were detected using an spatial ICA algorithm implemented in FieldTrip® Matlab software toolbox for MEG analysis [56]. Trials that included these segments were visually inspected and discarded.

Signals were estimated at ~550 equidistant points on the dorso-lateral aspect of the brain hull which we call *locations-of-interest* (LOIs). Their exact number varied across subjects in the range 545 – 574, depending on the brain anatomy of each individual subject. The mean distance between neighboring LOIs was 0.66 cm (SD 0.053). Fig 2a shows the LOIs over the brain surface of Subject 1. To reduce cross-talk between LOIs, amplitudes of the current dipoles were estimated following the octahedron method described in [57]. For more details about selection of LOI coordinates and source reconstruction see *Supplementary material*.

Since we aim to quantify how precisely-timed activity contributes to decoding, we used a point-process representation of the MEG signal as computed by [58]. Specifically, each analog signal from an MEG recording channel was transformed into a point process by detecting sharp activity transient which matched a template, corresponding to a half of a beta cycle that was previously observed with MEG, EEG and LFP in multiple brain regions [59–61]. Such transient events allow precise quantification of- time-lags. Here, detected event that fits the shape of interest is called mini evoked response (mini-ER). Importantly, previous works [58,63] have shown that the rate of transient beta events and the density map of mini-ERs across the brain allowed discriminating among cognitive states in evoked and ongoing neuronal activity. For more details about the conversion of the time series to point-process, see *Supplementary material*.

Fig 2 illustrates the data structure after preprocessing stages that is used for the analysis. Fig 2b shows a raster plot in which time-series from all 550 locations are transformed to a sparse representation of mini-ERs across time.

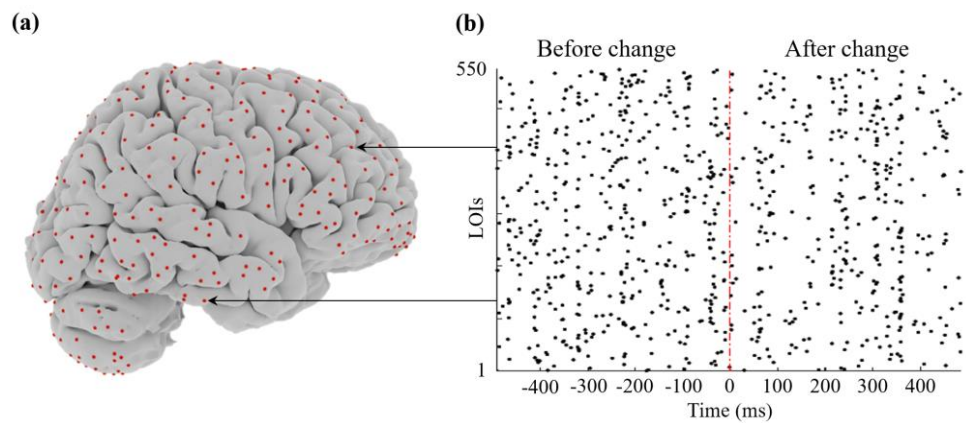

**Fig 2. Data structure. (a)** 550 locations of interest (LOIs, red dots) are shown across the surface of the cortex and cerebellum of Subject 1. **(b)** Brain-wide activity pattern: The activity recorded in a single trial of each condition across all 550 LOIs of Subject 1. Each row corresponds to an LOI, as illustrated by the arrows pointing to panel a. Time zero corresponds to the time of the first auditory stimulus that indicates a change in meter. Each point denotes a mini-ER event.

**Candidate decoding schemes**

We investigate three candidate properties of the neural activity and their possible role in decoding. Specifically, we calculate three summary statistics over sparse sequences of mini-ER events, each yielding a decoding scheme: (1) *Spatial decoding* – the distribution of activity across the brain. (2) *Spatio-temporal decoding* – the distribution of first-event latency across the brain. (3) *Time-difference decoding* – the distribution of pairwise time differences. To guarantee a meaningful comparison, the same preprocessing and decoding pipeline were applied for all three features. Below, we first discuss the three decoding schemes and then the decoding pipeline.

First, for *Spatial decoding*, we consider the spatial distribution of activity across LOIs. A natural idea is to use the rate of events occurring in each LOI to discriminate between conditions. Indeed, counting the mini-ERs across LOIs has been previously shown to discriminate well between high-level cognitive states [58]. This *spatial decoding* approach may be viewed as the parallel to spike-count decoding from multiple electrodes in spike-train analysis. Features were computed by counting the number of events occurring in each LOI for a single trial. This yielded a vector of 550 (number of LOIs) counts per trial.

Second, for *spatio-temporal decoding*, we consider the latency of the first mini-ER event from the stimulus onset for each LOI. In neural decoding from spike trains, the timing of the first spike after stimulus onset carries considerable information about the stimulus and is useful for decoding (e.g. 62–64). This approach allows representing a full-length point-process with a single value that is highly informative and can be estimated robustly. We computed the latency of the first mini-ER event from the onset of the latest stimulus for each LOI separately computed at 1-ms temporal resolution. This yields a vector of 550 latency values per trial.

Third, for *time-difference decoding*, we consider time differences between pairs of mini-ERs occurring in different LOIs. As discussed above, the behavioral task studied here may involve a cascade of responses in multiple brain areas. It is therefore natural to consider the role of relative time differences of activity events across pairs of brain regions for decoding. Time differences can be viewed as a generalization of the spatio-temporal model, but instead of considering delays compared to a common stimulus onset, we consider time differences between pairs of events. This is likely to capture better the fact that higher-order functions are not well locked to the stimulus but rather to their "preceding" processing stages. We define an *event-pair time difference* to be the time lag between the occurrence of events in two separate LOIs within a time window of 40ms. This time window length was selected based on the work of Tal and Abeles [28], however the analysis is not sensitive to this specific value, changing it to 50ms resulted in a reduction of up to 2.5% in accuracy. We extract all the time differences between mini-ER events in all pairs of LOIs as long as the time difference does not exceed 40ms. When multiple events were observed within a single LOI, all these events, with all their time differences, were used in training and inference. Below, each entry in a vector is treated as a statistic of the response. This approach yields a feature vector with 550*549/2=~160K time difference values per trial.

**Brain decoding from MEG signals**

We compare the classification accuracy of the three decoding schemes described above. To guarantee a meaningful comparison where the difference between decoding schemes is only attributed to differences in features, all three approaches were compared using a common decoding framework that aims to address two main challenges in MEG decoding: ***sample size***, and ***interpretability***. First the dimensionality of the data is very large in comparison to the number of samples, because only a small number of repeats can be collected while guaranteeing that a subject remains engaged and attentive to the task. The second challenge is that the decoding models we train should be interpretable to enable us to map the informational signals back to the brain.

Next, we describe the general framework of our approach for learning a decoder based on a given aspect of neural activity. Pseudocode for this pipeline is provided in the appendix.

**The decoding pipeline**

Our approach follows three steps for each given decoding scheme: (1) We rank and select candidate features by how discriminative each feature is in a training set; (2) We model the distributions of the selected features, so that we can estimate class likelihood for new trial from the test set. (3) We evaluate the accuracy of inference on a held-out test set.

**Step 1: Feature selection.** To achieve interpretable decoders, we select a small number of discriminative features by computing a saliency score for each feature independently and kept the *k* top features. Specifically, we compare the empirical distribution of the feature $S_k$ in the two conditions $p(S_k|C = before)$ and $p(S_k|C = after)$, and quantify their difference using a Mann-Whitney (MW) *p*-value. For robustness, we bootstrapped the estimate four times over the training trials, and used the mean *p*-value across the four folds to rank the features. We show below performance as a function of the number of features used for inference , while the other features are discarded. This allowed selecting a small number of LOIs that are highly discriminative about the condition.

**Step 2: Non-parametric model of the statistic distribution.** We next learn a non-parametric model of the distribution of each of the statistics in each condition, $\hat{p}(S_k|C)$. For *spatial decoding*, we use a maximum likelihood estimator of the multinomial, which means that $\hat{p}(S_k|C) = q$ is the fraction of trials that had occurrences. Fig 3d shows the distribution for the two conditions for the most discriminative LOI.

For *spatio-temporal decoding* and *time-difference decoding* the statistic is continuous hence we modelled $\hat{p}(S_k|C)$ by convolving the empirical distribution with a Gaussian kernel. The width of the Gaussian was chosen as a small constant multiplies by the standard deviation of $S_k$. We verified that the results were not sensitive to that kernel width by changing the constant up to an order of magnitude. We also added a uniform prior to avoid hard zero probability density, with a weight equal to a single extra trial, as if this prior is a single pseudo sample. Changing the weight of the prior distribution by up to 3 orders of magnitude only changed the accuracy by less than 3%. Fig 3e and 3f illustrates the resulting distributions for the case of spatio-temporal and time-difference decoding.

**Step 3: Inference.** We performed an optimal Bayesian decision under the assumption of conditional independence of features given the condition, and balanced classes. Specifically, at inference time, we classify a new trial based by computing the log-likelihood ratio under the assumption of conditional independence:

$$Score = \log\frac{\hat{p}(S_1,..S_k|C=1)}{\hat{p}(S_1,..S_k|C=2)} = \sum_{k=1}^{K} \log\frac{\hat{p}(S_k|C=1)}{\hat{p}(S_k|C=2)} \tag{1}$$

comparing this score to zero, and deciding C=1 if the score is positive and C=2 otherwise. Note that since the data is balanced, the log ratio of the prior-probabilities is zero, yielding the decision threshold to be zero.

**Evaluation**

We use 5-fold cross-validation procedure to evaluate all decoders on a common, randomly selected, held-out set (~36 trials per subject). We repeat the cross-validation procedure 25 times, to further estimate the variance of model predictions. All error rates reported below were averages over the five test sets used in this cross-validation procedure times the 25 repetitions. To further reduce the chance of over-optimistic evaluation (e.g. due to "graduate-student overfitting"), the data of half of the subjects was never used during exploration of the modeling parameters and was only used for the final evaluations after all the methods and design decisions have been set. For recent review about evaluation methods for decoding see [67].

**Cross-subject consistency analysis**

The decoding methods described above were trained and tested for each individual separately. The following section describes an approach to quantify which statistics are consistent across subjects. We focused on features of time-difference decoding because they significantly out-performed the other approaches in individual subjects.

We calculated a consistency score for each pair of anatomical regions to find region pairs that are consistently informative across subjects. The idea is to first map regions to a common ground, and then detect region pairs that are highly discriminative across all subject. Specifically, the consistency score was calculated in four steps. First, LOIs were mapped to the cortical surface and cerebellar surface of each individual subject. The surface representation of the cerebral and cerebellar cortex was derived from T1 MRI scans [68] and parcellated automatically to 84 regions $r_1, \ldots, r_{84}$ using public anatomical atlases [69,70]. This approach takes into account differences in brain anatomy across individuals, while allowing mapping across corresponding brain regions of multiple subjects. Second, we computed a discrimination scores for each location pair, being the minus log of the *p*-value of the Mann-Whitney U-test. Third, we mapped the discrimination scores using the parcellated brain surfaces to the anatomical regions, yielding a region-to-region matrix for each subject, $S_{subject} \in R^{84 \times 84}$. We normalized each matrix to have a zero mean and unit variance $E[S(i,j)] = 0, \text{Var}[S(i,j)] = 1$ where expectations were computed across all region-pairs.

Finally, based on these individual-subject matrices, we computed the Consistency Scores by taking the mean over all subjects, $S = \frac{1}{8}\sum_{subject=1}^{8} S_{subject}$. Since the S matrix has a zero mean and a unit variance, the value $S(i,j)$ is in fact a z-score for the region pair $\{r_i, r_j\}$. The score of a region pair is high if the corresponding LOI pair exhibit a discriminative pattern across multiple subjects, even though the actual pattern may change across subjects.

## 3. Results

There are two major challenges when developing brain decoding methods with human subjects. First, it is hard to keep people engaged in a task for a long period of time, and as a result, the number of samples tends to be very small for learning complex models. In our case, we wish to train decoders using a few dozen samples, while considering an input space whose dimensionality is in the 100,000s, hence standard regularized classifiers perform poorly. Second, decoding based on a large number of features is hard to interpret. The decoding methods we developed aim to address these two challenges, by applying aggressive feature selection based on non-parametric statistics, followed by standard inference. namely, making an optimal Bayesian decision. See the Methods section for full details.

We start by comparing the three decoding schemes described above: *spatial decoding, spatio-temporal decoding* and *time-difference decoding*. For each of these, we extracted statistics of the responses, and then conducted feature selection, modelling and inference as described in *Methods* (steps 1-3).

To illustrate discriminability of features, Fig 3 compares the empirical distribution (top) and the modeled distribution (bottom) of each of the three statistics we tested, for the most discriminative feature. (the most discriminative feature shown for illustration purposes only, all of the results described below achieved by considering all features unless stated otherwise). Specifically, Fig 3a shows the distribution of event counts in the two conditions for the LOI where these distributions were most easily discriminated. In this LOI, each single trial had between zero to two events occurring. This range of event counts was typical. As can be observed, the distribution of counts is hardly discriminative in this LOI.

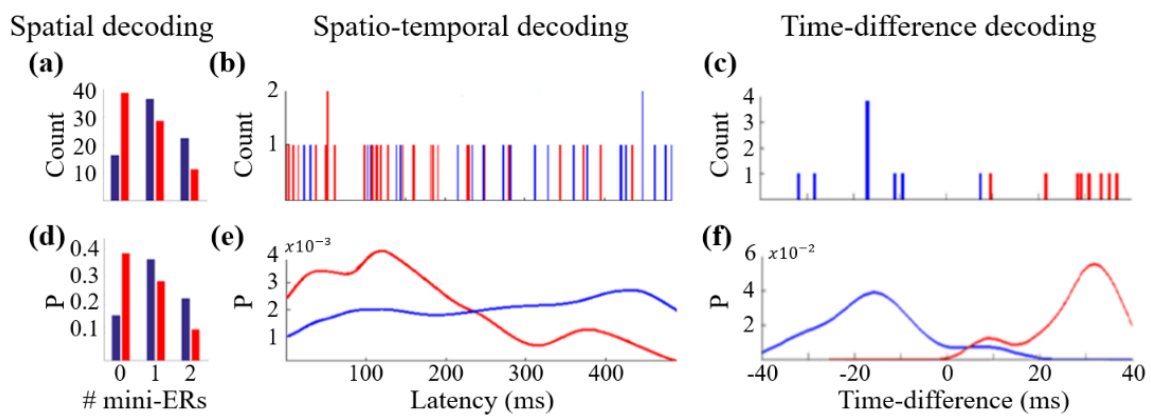

**Fig 3. Empirical and modeled distributions for spatial decoding (left), spatio-temporal decoding (middle) and time-difference decoding (right).** The top panel depicts the empirical distributions of the top predictive feature for each of the three decoding schemes. The bottom panel shows the modeled distributions for the same features, respectively. Red bars correspond to feature values from trials of the *before-change* condition, and blue bars to the *after-change* condition.

Fig 3b shows the empirical distribution of latencies in two conditions, again at the most discriminative LOI, and Fig 3e shows the modeled distributions for the same LOI (the model described in details in *Methods*). Fig 3c depicts the empirical and Fig 3f depicts the modeled distribution of event-pair time differences in the top-ranked pair of LOIs. In this best-LOI-pair example, the distribution of time-differences is very different in the two conditioning, making it relatively easy to infer which condition elicited the neural response. This can be seen by the small overlap between the blue and red distributions (Fig 3f). If classification was applied using this single feature, time-difference would clearly achieve superior accuracy.

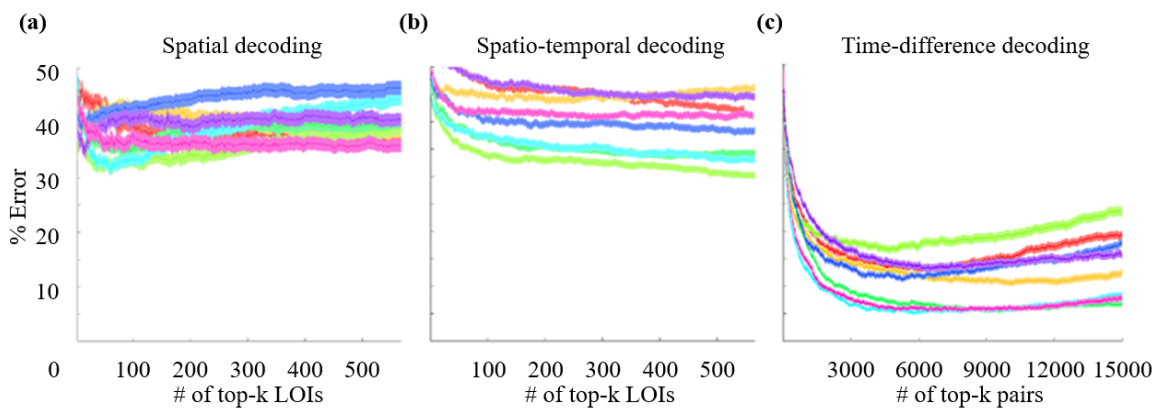

**Fig 4. Classification error rates achieved by the three decoding schemes.** Each curve shows the test-set error rate for a single subject, as a function of the number of features used for inference, selected based on the training set. Colors correspond to subject identity. The width of the curves corresponds to the standard error of the mean, over 125 values for each k value, computed by 25 bootstraps. **(a)** Decoding with spatial features. **(b)** Decoding with spatio-temporal features. **(c)** Decoding with time-differences.

Going beyond the best-sample illustration, Fig 4 compares the classification error for the three decoding schemes using many features. Each panel traces the error on the test set of trials, for eight subjects, as a function of the number of features selected using the training set. Fig 4a traces the error rate using spatial decoding. While the error is better than random (50%), it is quite poor overall, reaching an average accuracy of 62.37% across eight subjects (achieved with 58 LOIs). Fig 4b traces the error for spatio-temporal decoding, having error rates that are comparable to decoding using event counts and achieving a mean classification accuracy of 61.33%. Fig 4c traces the classification error for time-difference decoding, as a function of the number of pairs used for inference. Time-difference decoding strongly outperforms the other decoders, yielding a mean classification accuracy of 89.39% across 8 subjects (s.e.m. = 1.48), achieved with ~5000 pairs. Notably, the performance of the decoder decreases when lower-ranked pairs participate in the inference stage, suggesting that these lower-ranked pairs are spurious and do not provide information about Condition (before- or after- a change in rhythm). The superior accuracy of this model was consistent across subjects: an improvement was observed for each of the 8 subjects compared to spatial decoding and spatio-temporal decoding. This result supports the idea that multimodal behavioral tasks such as SMS involve sequences of activity events across multiple brain areas with temporal precision of a few milliseconds.

We next studied the spatial properties of time-difference decoding. We first tested how the length of a time lag may be related to the distance between its two corresponding LOIs. Fig 5a overlays the top 50 discriminative pairs over the reconstructed brain of Subject 1. The locations of these pairs are highly distributed across the brain. Interestingly, some distant pairs that are highly discriminative exhibit short time-lags (1-2 milliseconds). Clearly, they do not necessarily communicate directly, but rather, possibly both areas are activated by a common third region.

Fig 5b shows the joint distribution of distance in cm and absolute median time-lag across trials for Subject 1. The two are positively correlated (Spearman $p$-value= 4E-6, $n$=15000 most informative pairs), where pairs that are located far from each other have on average longer time lags. This is consistent with the basic notion that communication between distant brain regions takes longer than between neighboring ones. As a side note, the distribution of time lags peaks at 2ms and very few informative pairs have time lags above 30ms.

Fig 5c shows the strength of the correlation between the distance between LOIs within pair and the absolute median time lag for all eight subjects as a function of the number of LOI pairs taken. The correlation is statistically significant in six out of the eight subjects. For two subjects, the correlation is not statistically significant, but this may not be due to experimental noise, because inference accuracy in one of them is high.

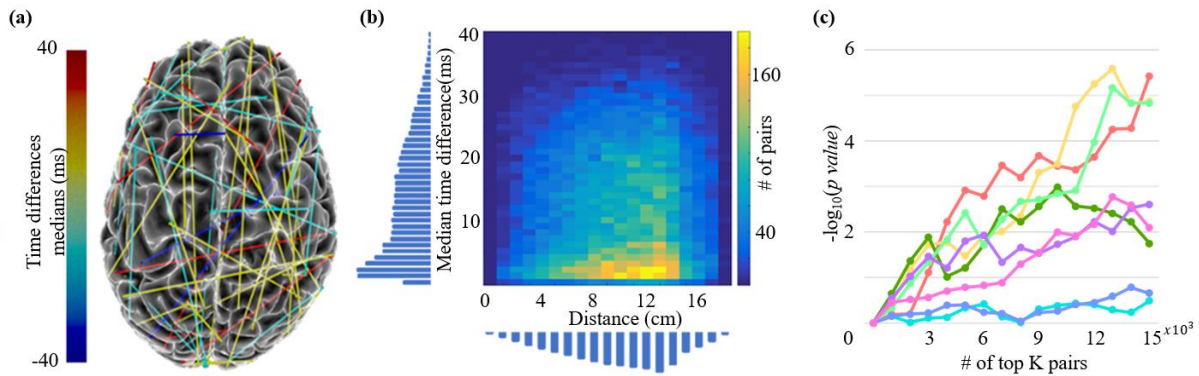

**Fig 5. Spatial properties of time-difference decoding. (a)** Brain-wide spatial distribution of most discriminative LOI pairs. Each line connects two LOIs that were among the top 50 most discriminative pairs, Line color corresponds to the median time differences in the before-change condition for Subject 1, overlaid on his reconstructed brain **(b)** Joint and marginal distributions of distances (in cm) and absolute value of median time-lag. Each discriminative pair contributed two values to the distribution, one median value per behavioral condition. Color correspond to the count in bin calculated over the 15,000 most discriminative pairs across all trials of Subject 1. **(c)** Spearman correlation significance between pair distance and the absolute median of time-lags within condition as a function of number of pairs used. The 8 curves present the results for 8 different subjects.

We further tested the robustness of *time-difference decoding* regarding the modeling parameters of the time-differences distributions, including the smoothing width, the weight of prior, and the maximum time-difference accounted by the model. We found the results to be robust across all these parameters, for instance, varying the Gaussian width by an order of magnitude (by multiplying or dividing by ~3) only changed the decoder performance by less than 2%. For details and figures see S1 Fig in the supplemental material.

**Cross-subject analysis**

The above analysis treated recordings from each subject separately, since neural activity varies across subjects. This variability has several sources, including anatomical and functional differences, and difference in the cognitive strategies that subjects applied to a task [71].

Beyond the individual patterns of activation, we seek to identify location pairs that are consistently informative across subjects. To discover such consistent pairs, we computed a cross-subject consistency score that quantifies which LOI pairs are consistently informative in multiple subjects. As a first step, we mapped LOIs into a unified set of 84 coarse regions (see Methods). We then computed, for each region pair, a region-based consistency score and its FDR -corrected significance $q$-value (see *Methods)*. Pairs of regions with $q$-value < 0.01 were treated as significantly consistent. Note that region pairs found using this procedure tend to repeat significantly across subjects, but the detailed distribution of time lags mapped to these regions does not necessarily repeat across subjects.

Fig 6 shows eleven pairs of regions that exhibited statistically significant cross subjects consistency scores ($q$-value<0.01). Interestingly, these consistent region pairs included regions known to be involved in error monitoring and error correction, including the insula, paracentral lobule, the cuneus and the cerebellum. As an example, the pair of regions that includes the primary auditory cortex (A1) and the insula is known to be involved in error awareness [72] and sensorimotor processing [73]. One may hypothesize that this pair captures a response to the deviant stimulus by A1, followed by detection of an incorrect tap by the insula.

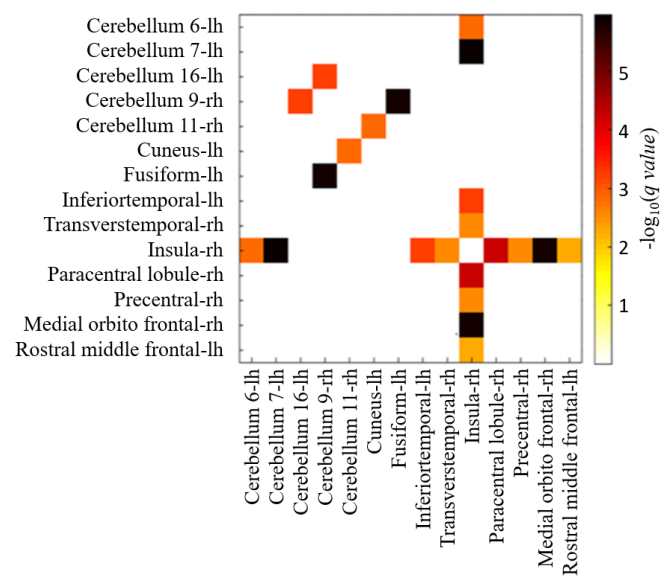

**Fig 6. Consistency of region pairs across subjects.** Shown are region pairs that are consistently discriminative across the 8 subjects. We tested all pairs of 84X84 regions defined on a common atlas [69,70]. The color code denotes the significance level as $-log_{10}(FDR\ q\ value)$. Only region pairs with $q < 0.01$ are colored. For example, the high value between the insula in the right hemisphere and region 7 of the left cerebellum indicates that in most of the subjects, pairs of LOIs located in these two regions show high divergence between time-difference distributions in the *before-change* and *after-change* conditions.

## 4. Discussion

We studied neural decoding of MEG signals in a rich auditory-motor task. To learn about which components of brain activity are predictive, we compared the performance of three decoding schemes, *spatial decoding*, *spatio-temporal decoding* and *time-difference decoding,* in the task of classifying single trials in a sensorimotor synchronization task. All three decoding schemes considered activity patterns across the whole brain but differed in how they use temporal relations in the signals. *Spatial decoding* ignored fine temporal structure, *spatio-temporal decoding* considered whole-brain spatio-temporal patterns, and *time-difference decoding* also used time differences between pairs of events across the brain. We discovered that *time-difference decoding* achieved significantly higher accuracy than *spatio-temporal decoding* and *spatial decoding*, reaching ~90% accuracy on held-out data at the single trial level. This high accuracy was robust over eight subjects, four of which were kept hidden while we developed the decoding methods.

The superior accuracy of decoding with time differences compared to decoding with onset time is consistent with a "cascade" model where a surge of activation events propagates across multiple regions. In such a model each station in the sequence accumulates timing noise, causing the activity patterns underlying compound multimodal behavior to be weakly locked to stimulus onset. It is important to highlight that highly-accurate decoding does not guarantee that the proposed feature is the one that the brain utilizes. However, one should interpret the accuracy of decoders as a lower bound on the amount of information carried by the type of decoding inspected.

The *time-difference decoding* approach analyzes each pair of brain locations separately, constructing a classifier that combines the "soft" voting of thousands of location pairs. Interestingly, these location pairs are not limited to a small number of brain regions, but virtually span the whole brain. Several effects may lead to this wide distribution. First, since the task in our experiment involved multimodal behavior, it may have activated multiple neural systems such as the auditory, motor, timing perception and error related systems together yielding many possible activity patterns. Second, the activity pattern involved in each of these sub-processes may also be distributed. For example, the motor system includes numerous regions including the premotor, primary and supplementary motor areas, the cerebellum and the basal ganglia. Third, it is possible that repeating the same behavior actually activates different pathways along the experiment.

We detected several region pairs that were consistently informative across subjects (Fig 5), even though informative LOI pairs were found independently in each subject. Many of these consistently-informative region pairs include areas that are known to be involved in sensorimotor synchronization (SMS). For example, the cerebellum has a documented role in timing during SMS tasks [74] and in error-correction [75]. Additionally, the insular cortex is considered to play a role in error awareness [72]. Originally, these regions were associated with SMS mainly using fMRI, where the temporal resolution is coarse. The results of this paper suggest that relevant information is coded in these regions not only in the *magnitude* of the activity, but also in the relative *timing* of their responses. Importantly, we did not limit the analysis to predefined regions of interest. As a result, we also find wide-spread activity events across many brain areas that were not traditionally associated with SMS. Interestingly, while these areas were predictive in single subjects, many of these areas were not consistently active across all subjects. One possible interpretation is that many activity patterns carry information at the level of individual subject, but due to inter-subject variability they are not observed if data is analyzed at the group level.

From the methodological point of view, the decoding task addressed here operates at an extremely "sample-starved" regime, where the dimensionality of the samples is two orders of magnitude larger than the number of samples. Such unfavorable sample-to-feature ratio is known to be very sensitive to overfitting, and we addressed this concern in several ways. First, we used a non-parametric approach to model the distribution of features within each condition, and carefully used cross validation to tune the hyper parameters of these approaches, Second, we were careful to strongly limit the number of hyper parameters, and used predefined values whenever possible. Third, we were also very concerned about a phenomenon sometimes called "graduate-student descent", where model accuracy improves through repeated trial-and-error iterations that leak information about the test data. To address this concern we kept half of the subjects as a held-out data and performed all method development on four subjects only, to avoid any informtion "leak". After model development, we found that the model performed equally well on subjects whose data was used during development and on subect data kept hidden. This is a strong evidence that the model complexity of our approach, in terms of the size of the hypothesis space, is well matched to the size of the data.

This study characterized properties of neural activity that enable decoding at the level of single trial within a subject, namely, predicting the underlying behavior from the activity of a *new trial* of the *same subject*. Another challenging task aims at decoding multimodal behavior for *new subjects*. In the cross-subject analysis described above we found eleven region pairs that were highly repeatable across subjects, but the majority of the neural activity that we found predictive, was not fully consistent across subjects. It remains an open challenge to understand this inter-subject variability and build models that can decode the brain activity of an unseen subject.

The peak accuracy using time difference decoding was achieved by combining information from thousands of LOI pairs, with each individual pair carrying only little information. Specifically, accuracy was highest when using 4000-5000 location pairs (varied across subjects). This large number of pairs was necessary: using the top hundred most informative location pairs caused a significant drop in accuracy, from ~90% to 64%. Such low information per pair suggests that the activity within each pair is highly variable. Unfortunately, it is hard to delineate what fraction of this variability is due to experimental noise and what fraction due to biological variability. By inspecting the distribution of time differences for various LOI pairs we find that the number of detected time differences for each LOI pair and condition is small: Only ~9% of the trials included co-activation in a given pair. This can be due to high experimental noise which hinders detecting events in the analogue MEG signal. Alternatively, it is possible that the effect is due to biological variability, for example if repeating the same behavior actually activates different pairs of regions. In such a case, only few trials will elicit events in the same region pairs.

This study highlights that a significant amount of information is carried by time differences of activations with milliseconds precision rather than by the magnitude of these activations in spatially distributed networks. Such time differences of activity events might be overlooked by an analysis that operates over a small number of local sources or analyses that involve coarse binning or averaging across trials. This emphasizes the importance of measuring neural activity with high temporal resolution even across distant brain locations.

## References

1. Milner PM. A model for visual shape recognition. Psychol Rev. American Psychological Association; 1974;81: 521.

2. Womelsdorf T, Schoffelen J-M, Oostenveld R, Singer W, Desimone R, Engel AK, et al. Modulation of Neuronal Interactions Through Neuronal Synchronization. Science (80- ). 2007;316: 1609–1612. doi:10.1126/science.1139597

3. Fries P. A mechanism for cognitive dynamics: Neuronal communication through neuronal coherence [Internet]. Trends in Cognitive Sciences. Elsevier Current Trends; 2005. pp. 474–480. doi:10.1016/j.tics.2005.08.011

4. Eckhorn R, Bauer R, Jordan W, Brosch M, Kruse W, Munk M, et al. Coherent oscillations: A mechanism of feature linking in the visual cortex? Biol Cybern. 1988;60: 121–130. doi:10.1007/BF00202899

5. Mioche L, Singer W. Chronic recordings from single sites of kitten striate cortex during experience-dependent modifications of receptive-field properties. J. Neurophysiol. 1989;62: 185–197.

6. Gray CM, Singer W. Stimulus-specific neuronal oscillations in orientation columns of cat visual cortex. Proc Natl Acad Sci. 1989;86: 1698–1702. doi:10.1073/pnas.86.5.1698

7. Gray CM, König P, Engel AK, Singer W. Oscillatory responses in cat visual cortex exhibit inter-columnar synchronization which reflects global stimulus properties. Nature. 1989;338: 334–7. doi:10.1038/338334a0

8. Gross J, Kujala J, Hamalainen M, Timmermann L, Schnitzler A, Salmelin R. Dynamic imaging of coherent sources: Studying neural interactions in the human brain. Proc Natl Acad Sci. 2001;98: 694–699. doi:10.1073/pnas.98.2.694

9. Pastalkova E, Itskov V, Amarasingham A, Buzsáki G. Internally generated cell assembly sequences in the rat hippocampus. Science. American Association for the Advancement of Science; 2008;321: 1322–7. doi:10.1126/science.1159775

10. Abeles M. Local Cortical Circuits An Electrophysiological Study [Internet]. Springer. 1982. Available: http://tocs.ulb.tu-darmstadt.de/21902720.pdf

11. Fries P. Rhythms for Cognition: Communication through Coherence. Neuron. NIH Public Access; 2015;88: 220–35. doi:10.1016/j.neuron.2015.09.034

12. Bastos AM, Vezoli J, Bosman CA, Schoffelen J-M, Oostenveld R, Dowdall JR, et al. Visual Areas Exert Feedforward and Feedback Influences through Distinct Frequency Channels. Neuron. 2015;85: 390–401. doi:10.1016/j.neuron.2014.12.018

13. Bosman CA, Schoffelen J-M, Brunet N, Oostenveld R, Bastos AM, Womelsdorf T, et al. Attentional stimulus selection through selective synchronization between monkey visual areas. Neuron. NIH Public Access; 2012;75: 875–88. doi:10.1016/j.neuron.2012.06.037

14. Gregoriou GG, Gotts SJ, Zhou H, Desimone R. High-frequency, long-range coupling between prefrontal and visual cortex during attention. Science. NIH Public Access; 2009;324: 1207–10. doi:10.1126/science.1171402

15. van der Meij R, Kahana M, Maris E. Phase-Amplitude Coupling in Human Electrocorticography Is Spatially Distributed and Phase Diverse. J. Neurosci. 2012;32: 111–123. doi:10.1523/JNEUROSCI.4816-11.2012

16. Bastos AM, Vezoli J, Fries P. Communication through coherence with inter-areal delays. Curr Opin Neurobiol. 2015;31: 173–180. doi:10.1016/j.conb.2014.11.001

17. Grothe I, Neitzel SD, Mandon S, Kreiter AK. Switching Neuronal Inputs by Differential Modulations of Gamma-Band Phase-Coherence. J. Neurosci. 2012;32: 16172–16180. doi:10.1523/JNEUROSCI.0890-12.2012

18. Jia X, Tanabe S, Kohn A. γ and the coordination of spiking activity in early visual cortex. Neuron. NIH Public Access; 2013;77: 762–74. doi:10.1016/j.neuron.2012.12.036

19. Zandvakili A, Kohn A. Coordinated Neuronal Activity Enhances Corticocortical

Communication. Neuron. NIH Public Access; 2015;87: 827–39. doi:10.1016/j.neuron.2015.07.026

20. Luczak A, McNaughton BL, Harris KD. Packet-based communication in the cortex. Nat Rev Neurosci. 2015;16: 745–755. doi:10.1038/nrn4026

21. Muller L, Reynaud A, Chavane F, Destexhe A. The stimulus-evoked population response in visual cortex of awake monkey is a propagating wave. Nat Commun. Nature Publishing Group; 2014;5: 3675. doi:10.1038/ncomms4675

22. Dima DC, Perry G, Singh KD. Spatial frequency supports the emergence of categorical representations in visual cortex during natural scene perception. Neuroimage. 2018;179: 102–116. doi:10.1016/j.neuroimage.2018.06.033

23. Goodale MA, Milner AD. Separate visual pathways for perception and action. Trends Neurosci. 1992;15: 20–5.

24. King J-R, Gramfort A, Schurger A, Naccache L, Dehaene S. Two Distinct Dynamic Modes Subtend the Detection of Unexpected Sounds. Kiebel S, editor. PLoS One. Public Library of Science; 2014;9: e85791. doi:10.1371/journal.pone.0085791

25. Pulvermuller F, Shtyrov Y. Spatiotemporal Signatures of Large-Scale Synfire Chains for Speech Processing as Revealed by MEG. Cereb Cortex. 2009;19: 79–88. doi:10.1093/cercor/bhn060

26. Repp BH, Su Y-H. Sensorimotor synchronization: a review of recent research (2006--2012). Psychon Bull Rev. Springer; 2013;20: 403–452.

27. Coull JT, Cheng R-K, Meck WH. Neuroanatomical and Neurochemical Substrates of Timing. Neuropsychopharmacology. Nature Publishing Group; 2011;36: 3–25. doi:10.1038/npp.2010.113

28. Tal I, Abeles M. Temporal accuracy of human cortico-cortical interactions. J. Neurophysiol. Am Physiological Soc; 2016;115: 1810–1820.

29. de-Wit L, Alexander D, Ekroll V, Wagemans J. Is neuroimaging measuring information in the brain? Psychon Bull Rev. Springer US; 2016;23: 1415–1428. doi:10.3758/s13423-016-1002-0

30. Kriegeskorte N, Goebel R, Bandettini P. Information-based functional brain mapping. Proc Natl Acad Sci. 2006;103: 3863–3868. doi:10.1073/pnas.0600244103

31. Quian Quiroga R, Panzeri S. Extracting information from neuronal populations: information theory and decoding approaches. Nat Rev Neurosci. Nature Publishing Group; 2009;10: 173–185. doi:10.1038/nrn2578

32. Besserve M, Jerbi K, Laurent F, Baillet S, Martinerie J, Garnero L. Classification methods for ongoing EEG and MEG signals. Biological Research. Sociedad de Biología de Chile; 2007. pp. 415–437. doi:10.4067/S0716-97602007000500005

33. Blankertz B, Lemm S, Treder M, Haufe S, Müller K-R. Single-trial analysis and classification of ERP components — A tutorial. Neuroimage. Academic Press;

2011;56: 814–825. doi:10.1016/J.NEUROIMAGE.2010.06.048

34. Perreau Guimaraes M, Wong DK, Uy ET, Grosenick L, Suppes P. Single-Trial Classification of MEG Recordings. IEEE Trans Biomed Eng. 2007;54: 436–443. doi:10.1109/TBME.2006.888824

35. Santana R, Bielza C, Larrañaga P. Regularized logistic regression and multiobjective variable selection for classifying MEG data. Biol Cybern. Springer-Verlag; 2012;106: 389–405. doi:10.1007/s00422-012-0506-6

36. Cecotti H, Barachant A, King JR, Sanchez Bornot J, Prasad G. Single-trial detection of event-related fields in MEG from the presentation of happy faces: Results of the Biomag 2016 data challenge. 2017 39th Annual International Conference of the IEEE Engineering in Medicine and Biology Society (EMBC). IEEE; 2017. pp. 4467–4470. doi:10.1109/EMBC.2017.8037848

37. Huttunen H, Manninen T, Kauppi J-P, Tohka J. Mind reading with regularized multinomial logistic regression. Mach Vis Appl. Springer Berlin Heidelberg; 2013;24: 1311–1325. doi:10.1007/s00138-012-0464-y

38. Dähne S, Meinecke FC, Haufe S, Höhne J, Tangermann M, Müller K-R, et al. SPoC: A novel framework for relating the amplitude of neuronal oscillations to behaviorally relevant parameters. Neuroimage. Academic Press; 2014;86: 111–122. doi:10.1016/J.NEUROIMAGE.2013.07.079

39. Bekhti Y, Zilber N, Pedregosa F, Ciuciu P, Wassenhove V Van, Gramfort A, et al. Decoding perceptual thresholds from MEG/EEG. Pattern Recoginition in Neuroimaging (PRNI). 2014; Available: https://hal.archives-ouvertes.fr/hal-01032909/document

40. van Gerven M, Hesse C, Jensen O, Heskes T. Interpreting single trial data using groupwise regularisation. Neuroimage. Academic Press; 2009;46: 665–676. doi:10.1016/J.NEUROIMAGE.2009.02.041

41. de Brecht M, Yamagishi N. Combining sparseness and smoothness improves classification accuracy and interpretability. Neuroimage. Academic Press; 2012;60: 1550–1561. doi:10.1016/J.NEUROIMAGE.2011.12.085

42. Kauppi J-P, Parkkonen L, Hari R, Hyvärinen A. Decoding magnetoencephalographic rhythmic activity using spectrospatial information. Neuroimage. Academic Press; 2013;83: 921–936. doi:10.1016/J.NEUROIMAGE.2013.07.026

43. Rieger JW, Reichert C, Gegenfurtner KR, Noesselt T, Braun C, Heinze H-J, et al. Predicting the recognition of natural scenes from single trial MEG recordings of brain activity. Neuroimage. Academic Press; 2008;42: 1056–1068. doi:10.1016/J.NEUROIMAGE.2008.06.014

44. Quandt F, Reichert C, Hinrichs H, Heinze HJ, Knight RT, Rieger JW. Single trial discrimination of individual finger movements on one hand: A combined MEG and EEG study. Neuroimage. Academic Press; 2012;59: 3316–3324.

doi:10.1016/J.NEUROIMAGE.2011.11.053

45. Bradberry TJ, Rong F, Contreras-Vidal JL. Decoding center-out hand velocity from MEG signals during visuomotor adaptation. Neuroimage. Academic Press; 2009;47: 1691–1700. doi:10.1016/J.NEUROIMAGE.2009.06.023

46. Akram S, Presacco A, Simon JZ, Shamma SA, Babadi B. Robust decoding of selective auditory attention from MEG in a competing-speaker environment via state-space modeling. Neuroimage. Elsevier; 2016;124: 906–917.

47. Kuo P-C, Chen Y-T, Chen Y-S, Chen L-F. Decoding the perception of endogenous pain from resting-state MEG. Neuroimage. Elsevier; 2017;144: 1–11.

48. Wardle SG, Kriegeskorte N, Grootswagers T, Khaligh-Razavi S-M, Carlson TA. Perceptual similarity of visual patterns predicts dynamic neural activation patterns measured with MEG. Neuroimage. Elsevier; 2016;132: 59–70.

49. Seeliger K, Fritsche M, Güçlü U, Schoenmakers S, Schoffelen J-M, Bosch SE, et al. Convolutional neural network-based encoding and decoding of visual object recognition in space and time. Neuroimage. Academic Press; 2017; doi:10.1016/J.NEUROIMAGE.2017.07.018

50. van de Nieuwenhuijzen ME, Backus AR, Bahramisharif A, Doeller CF, Jensen O, van Gerven MAJ. MEG-based decoding of the spatiotemporal dynamics of visual category perception. Neuroimage. Academic Press; 2013;83: 1063–1073. doi:10.1016/J.NEUROIMAGE.2013.07.075

51. Dima DC, Perry G, Messaritaki E, Zhang J, Singh KD. Spatiotemporal dynamics in human visual cortex rapidly encode the emotional content of faces. Hum Brain Mapp. Wiley-Blackwell; 2018; doi:10.1002/hbm.24226

52. Brandman T, Peelen M V. Interaction between Scene and Object Processing Revealed by Human fMRI and MEG Decoding. J Neurosci. Society for Neuroscience; 2017;37: 7700–7710. doi:10.1523/JNEUROSCI.0582-17.2017

53. Repp BH. Sensorimotor synchronization: a review of the tapping literature. Psychon Bull Rev. Springer; 2005;12: 969–992.

54. Blecher T, Tal I, Ben-Shachar M. White matter microstructural properties correlate with sensorimotor synchronization abilities. Neuroimage. Academic Press; 2016;138: 1–12. doi:10.1016/j.neuroimage.2016.05.022

55. Tal I, Abeles M. Cleaning MEG artifacts using external cues. J Neurosci Methods. Elsevier; 2013;217: 31–38.

56. Oostenveld R, Fries P, Maris E, Schoffelen JM. FieldTrip: Open source software for advanced analysis of MEG, EEG, and invasive electrophysiological data. Comput Intell Neurosci. 2011;2011. doi:10.1155/2011/156869

57. Lots Shapira I, Robinson SE, Abeles M. Extracting cortical current dipoles from MEG recordings. J Neurosci Methods. Elsevier; 2013;220: 190–196.

58. Abeles M. Revealing instances of coordination among multiple cortical areas. Biol

Cybern. Springer; 2014;108: 665–675. doi:https://doi.org/10.1007/s00422-013-0574-2

59. Jones SR. When brain rhythms aren't rhythmic: implication for their mechanisms and meaning. Curr Opin Neurobiol. Elsevier; 2016;40: 72–80.

60. Tal I, Abeles M. Imaging the Spatio-Temporal Dynamics of Cognitive Processes at High Temporal Resolution. Neural Comput. 2018;

61. Sherman MA, Lee S, Law R, Haegens S, Thorn CA, Hämäläinen MS, et al. Neural mechanisms of transient neocortical beta rhythms: Converging evidence from humans, computational modeling, monkeys, and mice. Proc Natl Acad Sci U S A. National Academy of Sciences; 2016;113: E4885-94. doi:10.1073/pnas.1604135113

62. Abeles M, Goldstein MH. Multispike train analysis. Proc IEEE. 1977;65: 762–773. doi:10.1109/PROC.1977.10559

63. Shin H, Law R, Tsutsui S, Moore CI, Jones SR. The rate of transient beta frequency events predicts behavior across tasks and species. doi:10.7554/eLife.29086.001

64. Middlebrooks JC, Xu L, Eddins AC, Green DM. Codes for sound-source location in nontonotopic auditory cortex. J Neurophysiol. Am Physiological Soc; 1998;80: 863–881.

65. Nelken I, Chechik G, Mrsic-Flogel TD, King AJ, Schnupp JWH. Encoding Stimulus Information by Spike Numbers and Mean Response Time in Primary Auditory Cortex. J Comput Neurosci. 2005;19: 199–221. Available: http://ai2-s2-pdfs.s3.amazonaws.com/878a/e07d50b3cd4f41bf4050057f67a54d495f64.pdf

66. Shamir M. The temporal winner-take-all readout. PLoS Comput Biol. Public Library of Science; 2009;5: e1000286.

67. Varoquaux G, Raamana PR, Engemann DA, Hoyos-Idrobo A, Schwartz Y, Thirion B. Assessing and tuning brain decoders: Cross-validation, caveats, and guidelines. Neuroimage. 2017;145: 166–179. doi:10.1016/j.neuroimage.2016.10.038

68. Dale AM, Fischl B, Sereno MI. Cortical Surface-Based Analysis. Neuroimage. 1999;9: 179–194. doi:10.1006/nimg.1998.0395

69. Buckner RL, Krienen FM, Castellanos A, Diaz JC, Yeo BTT. The organization of the human cerebellum estimated by intrinsic functional connectivity. J Neurophysiol. 2011;106: 2322–2345. doi:10.1152/jn.00339.2011

70. Desikan RS, Ségonne F, Fischl B, Quinn BT, Dickerson BC, Blacker D, et al. An automated labeling system for subdividing the human cerebral cortex on MRI scans into gyral based regions of interest. Neuroimage. Elsevier; 2006;31: 968–980.

71. Mechelli A, Penny WD, Price CJ, Gitelman DR, Friston KJ. Effective connectivity and intersubject variability: using a multisubject network to test differences and commonalities. Neuroimage. Elsevier; 2002;17: 1459–1469.

72. Klein TA, Ullsperger M, Danielmeier C. Error awareness and the insula: links to neurological and psychiatric diseases. Front Hum Neurosci. Frontiers Media SA; 2013;7: 14. doi:10.3389/fnhum.2013.00014

73. Kurth F, Zilles K, Fox PT, Laird AR, Eickhoff SB. A link between the systems: functional differentiation and integration within the human insula revealed by meta-analysis. Brain Struct Funct. Springer-Verlag; 2010;214: 519–534. doi:10.1007/s00429-010-0255-z

74. Molinari M, Leggio M, Thaut M. The cerebellum and neural networks for rhythmic sensorimotor synchronization in the human brain. The Cerebellum. 2007;6: 18–23. doi:10.1080/14734220601142886

75. Bijsterbosch JD, Lee K-H, Hunter MD, Tsoi DT, Lankappa S, Wilkinson ID, et al. The Role of the Cerebellum in Sub- and Supraliminal Error Correction during Sensorimotor Synchronization: Evidence from fMRI and TMS. J Cogn Neurosci. 2011;23: 1100–1112. doi:10.1162/jocn.2010.21506